\documentstyle[12pt]{article}
\begin{document}
\title{Non Linear Current Response of a Many-Level Tunneling System:
   Higher Harmonics Generation}
\author{Y.Goldin\footnote{goldin@bgumail.bgu.ac.il}
        and Y.Avishai\footnote{yshai@bgumail.bgu.ac.il}\\
   Physics Department,
   Ben Gurion University of the Negev, \\
   Beer Sheva, Israel}
\date{13 May 1996}
\maketitle
\begin{abstract}
The   fully  nonlinear  response  of  a  many-level  tunneling
system  to  a  strong  alternating  field  of  high  frequency
$\omega$   is   studied  in  terms  of  the  Schwinger-Keldysh
nonequilibrium  Green  functions. The nonlinear time dependent
tunneling   current  $I(t)$  is  calculated  exactly  and  its
resonance  structure  is  elucidated.  In  particular,  it  is
shown   that   under  certain  reasonable  conditions  on  the
physical   parameters,   the   Fourier  component  $I_{n}$  is
sharply  peaked  at $n=\frac {\Delta E} {\hbar \omega}$, where
$\Delta  E$  is the spacing between two levels. This frequency
multiplication  results  from  the highly nonlinear process of
$n$  photon  absorption (or emission) by the tunneling system.
It  is  also  conjectured  that  this  effect (which so far is
studied  mainly  in  the context of nonlinear optics) might be
experimentally feasible.
\end{abstract}
\begin{flushright}
Submitted to Phys.Rev.B\\
cond-mat/9605172
\end{flushright}

\newpage
\section{Introduction}
The physics of resonance tunneling through a single quantum well is
at the center of theoretical and experimental activity for more than
two decades. Electron (or hole) confinement between penetrable barriers
in a semiconductor enables the investigation of numerous important phenomena
such as negative differential resistance, Coulomb blockade, 
single electron tunneling, single electron pump
and many others. The pertinent physics is very attractive both because of
its richness and its potential device oriented nature.

At the core of the phenomena of resonance tunneling lies the relatively simple
picture of a small cavity connected by tunneling barriers to two
particle reservoirs (also termed as leads). The dimensions of the cavity
are small enough so that the energy levels inside it are well separated.
These levels might be either single particle levels whose spacing is
determined solely by geometrical considerations or a few particle levels
determined by interactions (as in the Coulomb blockade systems). 
If there is a difference between the chemical potential of the
left lead ($\mu_{L}$) and that of the right lead, ($\mu_{L}$)
 a tunneling current occurs between the two leads. If an energy level
of electrons in the cavity occurs between $\mu_{L}$ and $\mu_{R}$ this 
current displays a resonance structure. Evidently, the pertinent physics
is time independent, namely, one speaks here of a DC current. 
Moreover, in most cases, resonance tunneling
through a single level can be treated within the formalism of linear response.

Recently, interest is directed toward non linear time dependent transport
phenomena in double barrier resonance tunneling systems. Experimentally,
investigation of AC current in mesoscopic devices proves to be 
feasible \cite{Price94}. (Yet, theoretical analysis of the above 
experiment remains in the realm of linear response 
\cite{ButtPrTh93, Liu94,Pieper94}). 

The relevant physics inevitably becomes richer and more difficult to
analyze. It touches upon qualitatively new phenomena which depend on how
space and time dependent electronic states interfere. 
Among the effects which are
inherently based on non-linear response one might consider 
electron pumps \cite{Greelings90,Kouwenhoven91, Wingreen96},
lasers \cite{Faist94}, photon assisted tunneling \cite{TienGordon63,
KouwMcEuen-PRB, KouwMcEuen-PRL, BruScho94, Akiyama94, Blick} and
 others.

In the present work we concentrate on a relatively new effect,
namely, frequency multiplication of current response. As a
theoretical model one may consider a double barrier resonance tunneling
system containing at least two quantum levels whose spacing $\Delta$
introduces a new energy scale into the problem.
The system is then subject to a strong monochromatic
AC voltage of strength $W$ and frequency $\omega$, which results
in a non-linear current response $I(t)$, the main object of our
study. The combination of strong AC electric field and level 
interaction with the electrons in the reservoirs might lead to
transitions between these levels which are assisted by many photon
emission (or absorption), a highly non-linear effect. It can be
analyzed in terms of the Fourier components $I_{n}$ of the current
$I(t)$. In linear response, one expects $I_{0}$ and $I_{1}$ to be
the only non-zero components.  Here, however, higher components may
be significant. As we show below, $I_{n}$ as a function of $n$ is
peaked at $n=W/\Delta$, namely, there is a resonance when $n$ photons
are absorbed or emitted following transition between the two levels.
This kind of frequency multiplication is familiar in non-linear optics
but, to our best knowledge has not yet been investigated in micro-
electronics. We give below some realistic estimates of this effect.

As for the theoretical treatment of the above model, 
we start from the familiar tunneling Hamiltonian and employ
the Schwinger-Keldysh nonequilibrium
Green functions, through which the current is
 calculated in a straightforward
 way. In order to avoid complications while stressing
 the important role of two level quantum wells
we restrict ourselves in this work to non-interacting
particles. Principally, interaction effects can partially be 
included within screening or Hartree-Fock schemes.
 As we explain in the next section, the case of a Coulomb blockade system 
in which
the interaction is reflected through the charging energy is qualitatively 
included in the framework of our approach. The physics of non-linear AC 
response of a general interacting
system in a resonance tunneling device contains new effects which go beyond
the scope of the present work.

In the following section the problem is formulated, and the various
parts of the pertinent Hamiltonians are defined and justified. Then,
in section $3$ method of solution in terms of the Keldysh Green function
is introduced. In particular, the free particle Green functions are
written down and the Dyson equation for the ``lesser'' 
Green function is derived. Section $4$ is devoted to the discussion
of the tunneling current. An exact expression for the current is
 derived in terms of Keldysh Green functions and the origin of 
resonances in the non-linear response is explicitly  elucidated.
Analysis of higher harmonic generation is carried out in section $5$,
where numerical results are presented and the conditions for obtaining
peaks at higher harmonics are discussed. The paper is then
concluded with a short summary. In Appendix $A$ there is a proof
that a pure time dependent potential has no physically observable
effect, while some technical point related to the derivation of
Dyson equation in the Keldysh formalism is explained in Appendix $B$.

\section{Formulation of the problem}

We consider a structure where a charge carrier has two barriers in its way 
like the one drawn in fig.~\ref{dox-1}. The structure is analogous to the 
Fabry - Perot 
resonator -- the motion of a carrier is almost 
quantized in the central region but it still can escape into the leads. Thus 
the energy levels of the central region provide resonances for the 
transmission of the charge carries from one lead to another. Hereafter we 
will speak about electrons. The same type of structure can be made for holes 
but degeneracy of the valence band can complicate the calculation. 
An arbitrary combination of dc- and ac - potentials is applied to the 
structure (the potential differences are still required to be small compared 
with the Fermi energies). Our main goal is to 
 calculate the time - dependent currents in the system. 

Before doing this let us mention possible experimental realizations.
Practically the appropriate structure 
can be obtained in several ways:
\begin{enumerate}
  \item By putting several layers of different semiconductors having matching 
     lattices and different gap widths one on top of another 
     \cite{ChEsTsu74,Sollner83,Sollner84,Whitaker88,Goldman87-PR-L,
Goldman87-PR-B,Reed88,Rydberg89,Brown91,SuGoldCunn92,Dellow,Gueret, 
Ashoori92}. 
     The electrons then move in the direction perpendicular to the layers. 
     As an example, the profile of the conduction band for a particular case 
     of a {\em GaAs} - $\mbox{\em Al}_{x}\mbox{\em Ga}_{1-x}\mbox{\em As}$ 
     structure is shown in fig.~\ref{dox-1}.
  \item By application of an external gate potential 
     \cite{KouwMcEuen-PRB,KouwMcEuen-PRL,Meirav90,McEuenMWK91,FordPepper,
           FieldPepper93,FoxmanMWK93,
KlitzPloog93,MeurerPloog,JohnKouwenh92,
Nakazato,Heiblum,Molenkamp}. 
     The electrons then move along the layers but their movement is restricted 
     by the external repulsive potential. The gate is shaped such that the 
     electrons move in one direction and two barriers are formed in their way. 
  \item By placing an impurity in a tunnel barrier 
     \cite{Fowler86,Kopley88,RalphBuhr}. 
\end{enumerate} 

In order to describe the dynamics of the system we use the tunneling 
Hamiltonian method \cite{Bardeen,Cohen}. 
The total Hamiltonian is: 
\newcommand{\ek}{\epsilon_{k}}
\newcommand{\en}{\epsilon_{n}}
\newcommand{\ep}{\epsilon_{p}}
\newcommand{\ab}{a_{k}}
\newcommand{\ad}{a_{k}^{+}}
\newcommand{\cc}{c_{n}}
\newcommand{\cd}{c_{n}^{+}}
\newcommand{\bb}{b_{p}}
\newcommand{\bd}{b_{p}^{+}}
\begin{equation}
   H(t) = H_{f} + H_{ac}(t) + H_{T}                  \label{Ham-tot}
\end{equation}
where
\begin{equation} 
   H_{f} = \sum_{k} \ek \ad \ab + \sum_{n} \en \cd \cc   \label{Hf}
	  + \sum_{p} \ep \bd \bb
\end{equation}
is the Hamiltonian of free particles with neither ac-field nor coupling
between the leads and the central region. Here $\ek$, $\en$ and $\ep$ 
are free-particle energies
in the left lead, the central region and the right lead respectively.
The energy is taken 
with respect to the bottom of the conduction band in the central region.
The operators 
$\ad$, $\ab$, $\bd$, $\bb$, $\cd$, $\cc$ are creation and annihilation 
operators in the leads and in the central region. Furthermore, 
$k$ ($p$) are momenta in the direction perpendicular to the layers,
$n$ numbers the levels in the central region (hereafter $k$ ($p$) 
refer to the left (right) lead, $n$, $m$, $n'$, $m'$ -- to the central 
region), summation over $n$ in (\ref{Hf}) expresses the presence of more 
than one energy level. The third term,
\begin{equation} \label{Ham-transf}
   H_{T} = \sum_{k,n} \left( T_{nk}^{L} \cd \ab +
			     T_{nk}^{L \, *} \ad \cc \right)
	 + \sum_{p,n} \left( T_{pn}^{R} \bd \cc +
			     T_{pn}^{R \, *} \cd \bb \right)
\end{equation}
is the part responsible for the tunneling through the barriers, 
$T_{nk(p)}^{L(R)}$ are
transfer matrix elements between the leads and the central region.
Finally, the time dependent part is
\begin{equation}
   H_{ac}(t) = W_{L} \sin(\omega t) \cdot \sum_{k} \ad \ab  +
		W_{R} \sin(\omega t) \cdot \sum_{p} \bd \bb    \label{Ham-ac}
\end{equation}
where $ W_{L(R)} \sin(\omega t) $ are potential shifts of the leads with 
respect to the central region caused by an external ac-potential. 
dc - potential shifts are included into the energies $\epsilon_{k(n,p)}$. 
The arrangement in which
$W_{L}$ and $W_{R}$ have opposite signs corresponds to an application of 
an ac-bias as it was done in the experiments 
\cite{Sollner83,Rydberg89,MeurerPloog}. The choice 
$W_{L}=W_{R}$ describes an application of an ac-voltage to the gate electrode 
superimposed on the central region (see the experiments 
\cite{KouwMcEuen-PRB,KouwMcEuen-PRL}). 
We notice here that shifting both leads together (having the potential of 
the central region fixed) is equivalent to shifting the central region 
(having the potentials of the leads fixed) since an application of a uniform 
(time - dependent) potential is not observable even if it is arbitrary 
strong and arbitrary fast (see Appendix A). 
The situation $W_{L} = 0$, $W_{R} \neq 0$ corresponds to an
application of an ac - voltage only to one barrier as it was done in the 
experiment \cite{KouwMcEuen-PRB}. 

Our choice of $H_{ac}$ is based on the following assumptions:
\begin{enumerate}
  \item The electrons in the leads respond to an applied field very fast 
  since we deal with the frequencies much less than the plasma frequency. 
It means that any change of the external potential causes an immediate
rearrangement of the electrons. In other words,
the internal potential responds very quickly to an
external field \cite{ButtPrTh93,ChenTing91}.
  \item The concentration of the electrons in the leads is high enough to 
  screen an external field 
\cite{Frensley88,MW92,Sugimura93}. Therefore potential is uniform in the
leads and drops in the barriers 
\cite{TienGordon63,ChenTing91,Frensley88,Sugimura93,Tucker79,Johansson90,
Liu91,Cai91,SuYu91,Isawa92,Avishai,MW93:ac}.
   \item We used a widespread assumption 
   \cite{ChenTing91,Johansson90,Avishai,ChenTing90} that the probability of 
   direct transitions between the energy levels in the central region 
   due to the ac-field is small and can be neglected. 
\end{enumerate}

We do not restrict our consideration to the case of small $W_{L}$, $W_{R}$ 
(linear response). They can be arbitrary large. A strong ac-field 
($ W_{L} \gg \omega $ or $ W_{R} \gg \omega $) leads to a non - linearity.
One of its signatures is the
generation of current harmonics with frequencies much larger than $\omega$. 

The Hamiltonian (\ref{Ham-tot}) does not include Coulomb interaction. 
It is frequently omitted in the treatment of quantum wells 
\cite[and others]{ChenTing91,Frensley88,Sugimura93,Johansson90,Liu91,Cai91,
RiccoAzbel84,StAzLee85,WeilVinter87,JonsonGrincwaig87,Buttiker88,
Sok-JPC-88,SumFel88,Price92}. 
The number of electrons there is very large so the Coulomb interaction can 
be treated in the mean - field approximation. In fact it makes concentration - 
dependent corrections to the one - particle energies $\en$. It can lead to 
some changes in the dependence of the current on the bias (such as the 
appearance of a hysteresis 
\cite{Sollner83,SheardToombs88,Averin91,FiigJauho92}) 
or, may be, on the magnitude of the 
ac-field but the physics of the electron - photon interaction remains 
essentially within the independent particle model.
 In quantum dots the energy of the Coulomb 
interaction is much larger than the spacing between one - particle energy 
levels, level widths and temperature. The dc-bias, the magnitude of the 
ac-potentials ($W_{L}, W_{R}$) 
as well as the frequency are in general much smaller than the Coulomb 
interaction energy. Hence, to a  good approximation,
 every tunneling event involves only that electron which 
happens to have the highest energy. The Coulomb repulsion with the others 
only provides it with the potential energy needed to penetrate into the 
leads. Thus they serve as a background. We can forget (at least, qualitatively) 
about this background in our consideration of the interaction of that 
electron which provides the current with the ac-field. 

Rigorously speaking, the Hamiltonian (\ref{Ham-tot}) is one - dimensional. 
In QWs there are lateral degrees of freedom  provided by the motion along 
the layers. But at every particular value of the lateral momentum 
the problem is one - dimensional. Even if the dependence of electron energy 
on lateral momentum is important the total current can found by simple 
integration over it. The lateral degrees of freedom can be removed by a 
magnetic field perpendicular to the layers. 

\section{Solution in terms of Keldysh Green's functions.}

In order to work out the current we 
first have to analyze the electron propagation. To this end
we employ the non-equilibrium Green function 
technique suggested by Schwinger \cite{Schwinger61},
Kadanoff and Baym \cite{KadBaym62} and Keldysh \cite{Keldysh65} 
(for a review see Ref.%
\cite{SuYu91,LandauLif81,ChouSuHaoYu85,Mahan-b-90,Mahan87,RammerSm86}).
Since no equilibrium is required one can rigorously consider 
large perturbations and high
frequencies drawing the system far from its steady state.
The method uses  a time variable defined on two sides of the real axis.
It is equivalent to introduction two {\em independent} Green's functions, 
one which
characterizes the dynamical properties of the particles the other one
describes their distribution \cite{Keldysh65}. 

It is convenient to use the following two Green functions (the others can 
be expressed through them): 
\newcommand{\Gij}[1]{G^{#1}_{i,j}(t_{1},t)}
\begin{eqnarray}                    
  \Gij{r} & = & -i \theta(t_{1}-t) \left\langle
     \varphi_{i}(t_{1}) \varphi_{j}^{\dagger}(t)
 +  \varphi_{j}^{\dagger}(t) \varphi_{i}(t_{1})
				 \right\rangle    \label{Gij-def} \\
  \Gij{<} & = & i              \left\langle                     
		      \varphi_{j}^{\dagger}(t) \varphi_{i}(t_{1}) 
				 \right\rangle    \nonumber
\end{eqnarray}
where $ \varphi_{i}(t_{1}) \left( \varphi_{j}^{\dagger}(t) \right) $
are operators in the Heisenberg picture 
representing $a_{k}(t_{1})$, $c_{n}(t_{1})$, $b_{p}(t_{1})$
($a_{k}^{\dagger}(t)$, $c_{n}^{\dagger}(t)$, $b_{p}^{\dagger}(t)$)
in correspondence with what values the indexes
$i$ and $j$ take ($k$, $n$ or $p$). The first one is 
the usual retarded Green function. The advanced Green function
$\left( G_{ij}^{a} \right)$ is its Hermitian conjugate that is: 
$ \Gij{a} = \left[ G_{ji}^{r}(t,t_{1}) \right]^{*} $. The other one contains 
information about the distribution of electrons and their correlations. 
At $t_{1}=t$ it gives the one - particle density matrix ($\rho$) 
\cite{LandauLif81}: 
$ G_{ij}^{<}(t,t) = i N \rho_{ij}(t) $, where $N$ is the number of particles 
in the system. The advantage of the method is that 
it treats $G^{r}$ and  $G^{<}$ in a unified manner. Both of them can be found 
through the Dyson's equation \cite{Mahan-b-90}: 
\newcommand{\beq}{\begin{equation}}
\newcommand{\eneq}{\end{equation}}
\newcommand{\bea}{\begin{eqnarray}}
\newcommand{\enea}{\end{eqnarray}}
\bea
G^{r} & = & g^{r} \left[ 1 + \Sigma^{r} G^{r} \right]  \label{Dyson-r} \\
G^{<} & = & \left[ 1 + G^{r} \Sigma^{r} \right] g^{<}      \label{Dyson-<}
	    \left[ 1 + \Sigma^{a} G^{a} \right] + G^{r} \Sigma^{<} G^{a} 
\enea
Here multiplication implies summation (or integration) over space variables 
and integration over time, $g^{r}$ and $g^{<}$ 
are Green functions defined as 
(\ref{Gij-def}) but for an unperturbed Hamiltonian, 
$ \Sigma^{r} $, $\Sigma^{a}$ and $\Sigma^{<}$ are proper irreducible 
self - energies. 

An application of the formalism to tunneling systems is
especially powerful if one chooses $H_{T}$ as a perturbation
\cite{CarComNozSJ-1-71}).
Then then self-energies assume a very simple form:
\beq
\begin{array}{ccccccccl}                                    \label{Self en}
\Sigma_{kk'}^{r(a)} & = & \Sigma_{pp'}^{r(a)} & = &         
\Sigma_{kp}^{r(a)}  & = & \Sigma_{pk}^{r(a)}  & = & 0 \\
\Sigma_{nk}^{r} & = & \Sigma_{nk}^{a} & = &  T^{L}_{nk} & & & & \\  
\Sigma_{kn}^{r} & = & \Sigma_{kn}^{a} & = &  T^{L}_{kn} & = &   
					 T^{L \, *}_{nk} & & \\       
\Sigma_{np}^{r} & = & \Sigma_{np}^{a} & = &  T^{R}_{np} & & & & \\      
\Sigma_{pn}^{r} & = & \Sigma_{pn}^{a} & = &  T^{R}_{pn} & = &   
					 T^{R \, *}_{np} & & \\       
\Sigma^{<} & \equiv & 0 & & & & & & \\ 
\end{array}
\eneq
This is due to the fact that any vertex caused by $H_{T}$ has only two 
entries. The only diagram contributing to an irreducible self - energy is 
the simplest one drawn in fig.~\ref{tunn-diagrams}a. The next one 
drawn in fig.~\ref{tunn-diagrams}b is already reducible. 

In the next section we will show that tunneling currents can be expressed 
solely in terms of the Green functions in the central region. To find them 
we iterate (\ref{Dyson-r}) and get: 
\bea                                                
  G_{nm}^{r}(t_{1},t) & = & g_{nm}^{r}(t_{1},t) +  \label{eq-centr-r}   \\
		   & + & \sum_{n'} 
   \int_{-\infty}^{+\infty} \!\! \int_{-\infty}^{+\infty} \! \! dt_{2}  dt_{3}
       \, g_{nn}^{r}(t_{1},t_{2}) X_{nn'}(t_{2},t_{3}) G_{n'm}^{r}(t_{3},t)  
							       \nonumber  
\enea
\bea
  X_{nn'}(t_{2},t_{3}) & \equiv & X_{nn'}^{L}(t_{2},t_{3}) + 
				  X_{nn'}^{R}(t_{2},t_{3})     \label{X}  \\
  X_{nn'}^{L}(t_{2},t_{3}) & \equiv & 
	\sum_{k} T^{L}_{nk} g_{kk}^{r}(t_{2},t_{3}) T^{L}_{kn'}  \nonumber \\
  X_{nn'}^{R}(t_{2},t_{3}) & \equiv & 
	\sum_{p} T^{R}_{np} g_{pp}^{r}(t_{2},t_{3}) T^{R}_{pn'}    \nonumber
\enea

Equation (\ref{Dyson-<})-(\ref{Self en}) 
establish an algorithm for calculating $G_{nm}^{<}$ without an 
iteration but nevertheless it contains eleven terms on the right hand side. 
We showed (see Appendix B) that only two of them remain after a long enough 
period of time has passed from the moment the tunneling was switched on: 
\beq                                                  \label{eq-centr-<}
  G_{nm}^{<}(t_{1},t)  =   \sum_{n'm'} 
   \int_{-\infty}^{+\infty} \!\! \int_{-\infty}^{+\infty} \! \! dt_{2}  dt_{3}
       \, G_{nn'}^{r}(t_{1},t_{2}) Y_{n'm'}(t_{2},t_{3})
 G_{m'm}^{a}(t_{3},t),
\eneq
where
\bea
  Y_{n'm'}(t_{2},t_{3}) & \equiv & Y_{n'm'}^{L}(t_{2},t_{3}) + 
				   Y_{n'm'}^{R}(t_{2},t_{3})   \label{Y}  \\
  Y_{n'm'}^{L}(t_{2},t_{3}) & \equiv & 
	\sum_{k} T^{L}_{n'k} g_{kk}^{<}(t_{2},t_{3}) T^{L}_{km'}  \nonumber \\
  Y_{n'm'}^{R}(t_{2},t_{3}) & \equiv & 
	\sum_{p} T^{R}_{n'p} g_{pp}^{<}(t_{2},t_{3}) T^{R}_{pm'}    \nonumber
\enea
We notice that $G_{nm}^{<}$ does not depend on the initial distribution of 
electrons in the central region (because $g_{nm}^{<}$ do not appear in 
formula (\ref{eq-centr-<})). It depends only on their distributions in the 
leads through $g^{<}_{kk}$ and $g^{<}_{pp}$. 

It is easy to show (see Appendix A) that even under a time-dependent 
potential of an arbitrary large amplitude and an arbitrary high frequency 
the Green functions for electrons in isolated leads are given by 
``adiabatic - like'' expressions:
\bea                                                   \label{leadsGF}
  g^{r}_{k k'}(t_{2},t_{3}) & = & -i \delta_{k k'} \Theta(t_{2}-t_{3}) 
		    e^{-i \int^{t_{2}}_{t_{3}} 
		       \left[ H_{f}+H_{ac}(t) \right]_{kk} \,dt } = \\
		     & = & -i \delta_{k k'} \Theta(t_{2}-t_{3}) 
		    e^{-i \epsilon_{k}(t_{2}-t_{3}) + i \frac{W_{L}}{\omega} 
		 \left(\cos (\omega t_{2}) - \cos (\omega t_{3}) \right) } 
								\nonumber \\
  g^{<}_{k k'}(t_{2},t_{3}) & = & i \delta_{k k'} f_{L}(\epsilon_{k})  
		    e^{-i \int^{t_{2}}_{t_{3}} 
		       \left[ H_{f}+H_{ac}(t) \right]_{kk} \,dt } = 
								\nonumber \\
		     & = & i \delta_{k k'} f_{L}(\epsilon_{k})  
		    e^{-i \epsilon_{k}(t_{2}-t_{3}) + i \frac{W_{L}}{\omega} 
		 \left(\cos (\omega t_{2}) - \cos (\omega t_{3}) \right) } 
								   \nonumber
\enea               
provided that the time - dependent perturbing potential ($H_{ac}$) is 
uniform in every lead. Here $[\ldots ]_{kk}$ denotes matrix element, 
$ f_{L}(\ek) = \frac{1}{ e^{ \frac{\ek - \mu_{L}}{k \Theta} } + 1 } $ is the 
Fermi function for the left lead, $\mu_{L}$ is its chemical potential, 
$\Theta$ is the temperature. 
Formulas for $g^{r}_{pp'}$, $g^{<}_{pp'}$ can be obtained from 
(\ref{leadsGF}) by replacements $k, k' \rightarrow p, p'$, 
$ W_{L} \rightarrow W_{R} $, $ \mu_{L} \rightarrow \mu_{R} $. 

Now we substitute (\ref{leadsGF}) into (\ref{X}) and (\ref{Y}) and replace
the sum over $k$, $p$ by an integral over energy.
Further, we employ the ``wide - band 
approximation'' \cite{Wingreen89,LangrethNor91} assuming the elastic 
coupling to the leads 
\beq                                                        \label{Gamma}
  \Gamma_{nn'}^{L} \equiv 2 \pi \rho(\ek) T_{nk}^{L} T_{kn'}^{L} 
  \hspace{5mm} \mbox{and} \hspace{5mm} 
  \Gamma_{nn'}^{R} \equiv 2 \pi \rho(\ep) T_{np}^{R} T_{pn'}^{R}
\eneq
to be energy independent (i.e. they do not depend on $k$, $\ek$) 
within the band. The band widths and the Fermi energies are large compare to 
$W_{L}$, $W_{R}$. Thus (\ref{X}) and (\ref{Y}) take a rather simple form: 
\newcommand{\tone}{t_{1}}
\newcommand{\ttwo}{t_{2}}
\bea
  X_{nn'}(t_{1},t_{2}) & = & -\frac{i}{2} \Gamma_{nn'} \delta(t_{1}-t_{2}) 
							      \label{Xnn}  \\
  Y_{n'm'}(t_{1},t_{2}) & = & i \Gamma_{n'm'}^{L}             \label{Ynm}
	    \int_{V_{L}}^{+\infty} \! \! \frac{d\ek}{2\pi} f_{L}(\ek)
		 e^{-i \ek(t_{1}-t_{2}) + i \frac{W_{L}}{\omega} 
       	    \left(\cos (\omega \tone) - \cos (\omega \ttwo) \right) } \\ 
			& + & i \Gamma_{n'm'}^{R}             
	    \int_{V_{R}}^{+\infty} \! \! \frac{d\ep}{2\pi} f_{R}(\ep)
		 e^{-i \ep(t_{1}-t_{2}) + i \frac{W_{R}}{\omega} 
		    \left(\cos (\omega \tone) - \cos (\omega \ttwo) \right) }  
							      \nonumber 
\enea
Hereafter $V_{L}$ ($V_{R}$) are energies of the conduction band bottoms in 
the left (right) lead, 
$\Gamma_{nn'} \equiv \Gamma_{nn'}^{L} + \Gamma_{nn'}^{R} $. 
We can not put $ V_{L}, V_{R} \rightarrow - \infty $ in the formula (\ref{Ynm}) 
because it would lead to a divergence in the ac-current. 

We want to emphasize that substituting 
(\ref{Xnn}) into (\ref{eq-centr-r}) 
yields a time - independent equation. 
The problem then reduces to the task of finding the 
usual retarded Green function for a time - independent
 potential well with finite barriers.
This is due to the energy independence 
of $\Gamma_{nn'}^{L(R)}$. The Fourier transform of the equation is: 
\beq
  G_{nm}^{r}(\epsilon) + \frac{i}{2} 
     g_{nn}^{r}(\epsilon) \sum_{n'} \Gamma_{nn'} G_{n'm}^{r}(\epsilon) 
				=  g_{nm}^{r}(\epsilon)   \label{D-Ftrans} 
\eneq 
A rigorous solution for a two - level system 
shows that 
the mixing terms $G_{nm}^{r}, n \neq m $ give a contribution to the current 
of the order of $\Gamma / \left|\epsilon_{n} - \epsilon_{m} \right| $. 
We assume $ \Gamma \ll \left|\epsilon_{n} - \epsilon_{m} \right| $ and 
drop them out hereafter. With the same accuracy $G_{nn}^{r}$ are given by: 
\beq                                                   \label{GF-sol-r}
  G_{nn}^{r}(\epsilon) = \frac{1}{ \epsilon - \en + i \frac{\Gamma_{n}}{2} } 
\eneq
where                 
$\Gamma_{n} \equiv \Gamma_{nn} + \Gamma_{n}^{\mbox{\scriptsize in}} $, 
$ \Gamma_{n}^{\mbox{\scriptsize in}} $ are intrinsic level widths due to 
inelastic interactions, leakage of electrons into lateral directions etc. 
This is equivalent to the widely - used assumption that the quasilevels 
of a potential well with finite barriers have complex energies 
$ \en - i \frac{\Gamma_{n}}{2} $.
We see that the main contribution of tunneling 
in the {\em dynamical} properties is level broadening. 

Substituting (\ref{Ynm}) into (\ref{eq-centr-<}) gives an exact expression 
for $G^{<}$ as soon as we know $G^{r}$: 
{\samepage
\bea 
  G_{nm}^{<}(t_{1},t) & = &                              
    \int_{V_{L}}^{+\infty} \frac{d\ek}{2\pi}\,f_{L}(\ek) e^{-i \ek(t_{1}-t)}
	      \sum_{s,q = -\infty}^{\infty} i^{s-q+1} 
		J_{s}(\frac{W_{L}}{\omega}) J_{q}(\frac{W_{L}}{\omega}) 
	   e^{-i s \omega t_{1} + i q \omega t}  \cdot        \nonumber  \\
    & & \hspace{3.5cm} \cdot
	      \sum_{n'm'} \Gamma_{n'm'}^{L} G_{nn'}^{r}(\ek + s\omega) 
					   G_{mm'}^{r\,*}(\ek + q\omega) +
      						 \label{GF-sol-<}  \\
& + & \int_{V_{R}}^{+\infty} \frac{d\ep}{2\pi}\,f_{R}(\ep) e^{-i \ek(t_{1}-t)}
	      \sum_{s,q = -\infty}^{\infty} i^{s-q+1} 
		J_{s}(\frac{W_{R}}{\omega}) J_{q}(\frac{W_{R}}{\omega}) 
	e^{-i s \omega t_{1} + i q \omega t}  \cdot        \nonumber  \\
    & & \hspace{3.5cm} \cdot
	      \sum_{n'm'} \Gamma_{n'm'}^{R} G_{nn'}^{r}(\ep + s\omega) 
					   G_{mm'}^{r\,*}(\ep + q\omega)
								  \nonumber
\enea
}
We notice that the transition 
amplitudes $G_{nm}^{<}, n \neq m $ are important 
as we show in the next section. 

This difference between $G_{n m}^{r}$ and $G_{n m}^{<}$, $n \neq m$ arises 
from the difference between $X$ and $Y$ in the formulas 
(\ref{Xnn}), (\ref{Ynm}). The quantity
$X$ has such a simple form because $f(\epsilon_{k(p)})$ does not appear in 
$g_{k k}^{r}$, $g_{p p}^{r}$ so the summation over $k(p)$ (equivalent to 
integration over $\ek$($\ep$)) in formula (\ref{X}) gives a $\delta$ - 
function $\delta (t_{1} - t_{2}) $ if the band (i.e the range of 
$\epsilon_{k(p)}$) is wide. The occurrence of $f(\epsilon_{k(p)})$ in 
$g_{k k}^{<}$ ($g_{p p}^{<}$) leads to an upper limit of integration in 
(\ref{GF-sol-<}) around $\mu_{L(R)}$. If $\mu_{L(R)}$ is large enough the 
$\delta (t_{1} - t_{2}) $ is recovered. Then the whole solution becomes
invariant under time translation and the transition
amplitudes $G_{n m}^{<}$, 
$n \neq m$ are small, of the order of $ \Gamma /|\en - \epsilon_{m} | $. 

\section{The current} 

A double barrier structure is integrated within a circuit. The measured 
current is determined by its influence on the circuit. If the barriers are 
modeled as capacitive - resistant elements \cite{IngoldNaz} 
the currents in the leads are given by \cite{BruScho94}: 
\bea 
   I_{L}(t) & = &                                     \label{lead_curr} 
       \frac{C_{R} + C_{g}}{C} C_{L} \omega W_{L} \cos (\omega t) 
	    - \frac{C_{L}}{C} C_{R} \omega W_{R} \cos (\omega t)   +     \\ 
	    & + & 
       \frac{C_{R}+C_{g}}{C} I_{L}^{T}(t) - \frac{C_{L}}{C} I_{R}^{T}(t)  
								 \nonumber 
\enea 
where $I_{L}$ is the total current in the left lead $I_{L}^{T}(t)$ and 
$I_{R}^{T}(t)$ are tunneling currents through the left and right barriers 
respectively, $C_{L(R)}$ and $C_{g}$ are capacitances of the central region 
relative to the left (right) lead and a gate electrode (or another 
background), $C \equiv C_{L} +C_{R} +C_{g}$. The current in the right lead 
is obtained by replacing $L$ by $R$. The first two terms describe 
the contribution of capacitive currents (due to the presence of accumulation 
and depletion layers). They have the frequency of the input ac - voltage. But 
the tunneling currents contain also 
higher harmonics as we show below. 

The tunneling current from the left lead into the central region is defined 
by the change in the number of electrons in that lead: 
\beq                                                    \label{Curr-df} 
\begin{array}{rcccl}
 I_{L}^{T}(t) & = & e \left\langle \frac{d\hat{N}_{L}}{dt} \right\rangle 
	      & = & -\frac{i e}{\hbar} \sum_{km}
		    \left( T_{mk}^{L} \langle c_{m}^{+} \ab \rangle -
			 T_{mk}^{L \, *} \langle \ad c_{m} \rangle \right) \\
	& &   & = & -\frac{2 e}{\hbar} \mbox{Re} \left[ \sum_{km}
				T_{mk}^{L} G_{km}^{<}(t,t) \right]  \nonumber 
\end{array}
\eneq
where $\hat{N}_{L} \equiv \sum_{k} \ad \ab $. The tunneling current from the 
right lead into the central region ($I_{R}^{T}$) is obtained from the above 
expression by a change $ L \rightarrow R $, $ k \rightarrow p $, 
$ \ab \rightarrow \bb $. 

Using Dyson's equation for $G_{kn}^{<}$ allows one 
to express the tunneling currents 
through Green functions in the central 
region (which we found in the previous section): 
\beq                                                     \label{curr-gen}
 I_{L(R)}^{T}(t) = -\frac{2 e}{\hbar} \mbox{Re} 
		 \left\{ \sum_{nm}
   \int_{-\infty}^{+\infty} \! \! dt_{1} \, 
	  \left[ X_{mn}^{L(R)}(t,t_{1}) G_{nm}^{<}(t_{1},t)  +  
		 Y_{mn}^{L(R)}(t,t_{1}) G_{nm}^{a}(t_{1},t) \right]  \right\}  
\eneq

Under the assumption of energy independent coupling to the leads which we 
used in the previous section (see (\ref{Gamma})) we can use the formulas 
(\ref{Xnn}), (\ref{Ynm}), (\ref{GF-sol-r}) and (\ref{GF-sol-<}). 
Dropping out the terms that are always small of the order of 
$\Gamma / \left|\epsilon_{j} - \epsilon_{j'} \right| $ in comparison with 
the others we obtain after some algebra: 
\bea
  I_{L}^{T}(t) & = & \frac{1}{2} I_{L}^{0} +                \label{curr-fnl} 
		     \sum_{n=1}^{+\infty} \left| I_{L}^{n} \right| 
					 \cos (n\omega t + \phi_{L}^{n}), 
                        \mbox{where}  \phi_{L}^{n} = \arg I_{L}^{n}   \\
  I_{L}^{n} & = & \frac{2 e}{\hbar} \sum_{j} 
	\left[ A_{j}(n) + A_{j}^{*}(-n) + B_{j}(n) + B_{j}^{*}(-n) \right] 
								\nonumber  \\
  A_{j}(n) & = & i^{-n} \left[ 
		    i \Gamma_{jj}^{L} + \sum_{j'} 
			 \frac{ \left| \Gamma_{jj'}^{L} \right|^{2} } 
			      {\epsilon_{j'} - \epsilon_{j} - n \omega 
			       +i \frac{ \Gamma_{j} + \Gamma_{j'} }{ 2 } } 
			    \right] \cdot                        \nonumber \\
	   & & \cdot \sum_{s = -\infty}^{+\infty}  
		   J_{s}(\frac{W_{L}}{\omega}) J_{s+n}(\frac{W_{L}}{\omega}) 
	F \left( \frac{\epsilon_{j} - \mu_{L}}{\omega} - s , 
		 \frac{\mu_{L} - V_{L}}{\omega}, 
		 \frac{\Gamma_{j}}{\omega}, \frac{k \Theta}{\omega} \right)  
								 \nonumber \\
  B_{j}(n) & = & i^{-n} \sum_{j'} 
			 \frac{ \Gamma_{jj'}^{L} \Gamma_{j'j}^{R} } 
			      {\epsilon_{j'} - \epsilon_{j} - n \omega 
			       +i \frac{ \Gamma_{j} + \Gamma_{j'} }{ 2 } } 
				    \cdot                        \nonumber \\
	   & & \cdot \sum_{s = -\infty}^{+\infty}  
		   J_{s}(\frac{W_{R}}{\omega}) J_{s+n}(\frac{W_{R}}{\omega}) 
	F \left( \frac{\epsilon_{j} - \mu_{R}}{\omega} - s , 
		 \frac{\mu_{R} - V_{R}}{\omega}, 
		 \frac{\Gamma_{j}}{\omega}, \frac{k \Theta}{\omega} \right)  
								 \nonumber 
\enea
\[
  F \left( \frac{\epsilon_{j} - \mu}{\omega} - s , \frac{\mu - V}{\omega}, 
	   \frac{\Gamma_{j}}{\omega}, \frac{k \Theta}{\omega} \right)  
      =    \frac{1}{2\pi} \int_{V}^{+\infty} \, 
	   \frac{ f(\epsilon) d\epsilon }
	    { \epsilon + s \omega - \epsilon_{j} + i \frac{\Gamma_{j}}{2} },\;
      f(\epsilon) = \frac{1}{ e^{ \frac{\epsilon - \mu}{k \Theta} } + 1 }
\]
Analogous expressions hold for $I_{R}^{T}(t)$ as well. 
 For the rest of our discussion we choose to study $I_{L}^{T}$. 
We point out that the current consists of many harmonics with the frequencies 
$n\omega$. The sums over $j, j'$ express summation over all energy levels in 
the central region. We notice that the 
terms $A_{j}(n)$ and $B_{j}(n)$ are quite similar in structure. 
The first one includes the currents due to direct photon - assisted 
transitions between the left lead and the central region and due to the 
effect of the left lead on the population of the central region. 
The second one describes the current through the left barrier due to the 
influence of the right lead  on the electrons 
in the central region. The main difference between them (beside a few
replacements of L into R) is that 
the first term in the brackets 
in $A$ (i.e. $ i \Gamma^{L}_{j j} $) is absent in $B_{j}(n)$. 
At zero temperature the integral in the expression for $F$ can be easily 
computed: 
\[ 
  F \left( \frac{\epsilon_{j} - \mu}{\omega} - s , \frac{\mu - V}{\omega}, 
	   \frac{\Gamma_{j}}{\omega}, 0 \right)  
       =  \frac{1}{2\pi}  
	  \ln \frac{ \epsilon_{j} - s\omega - \mu -i \frac{\Gamma_{j}}{2} }
		   { \epsilon_{j} - s\omega -   V -i \frac{\Gamma_{j}}{2} } 
\] 
The rest of our calculation is done at zero temperature. 

The term in A containing $\Gamma_{jj}$ gives an additive contribution of 
different energy levels to the current. The same is true for the terms of 
the sums over $j'$ (in $A_{j}(n)$ and $B_{j}(n)$) with $j'=j$. 
The other terms in these sums 
are caused by the presence of different energy levels {\em together}. 
Every one is caused by a certain {\em pair} of energy levels. It is obvious 
that they have a resonance nature. The $j'$-th term can be large when 
$ \left| \epsilon_{j'} - \epsilon_{j} - n\omega \right| < 
					( \Gamma_{j'} + \Gamma_{j} ) / 2 $. 
Thus it can give large ac-current at 
$ n_{\mbox{\scriptsize res}}  \approx
		    \left| \epsilon_{j'} - \epsilon_{j} \right| / \omega $ 
(under condition $ W > \left| \epsilon_{j'}-\epsilon_{j} \right| $). 
If $ \omega \ll \left| \epsilon_{j'}-\epsilon_{j} \right| $ one has 
$ n_{\mbox{\scriptsize res}} \gg 1 $, i.e. the frequency in the output is 
much larger than in the 
input. We notice that this term  generates the 
$ n_{\mbox{\scriptsize res}} $-th 
harmonic exclusively. The others are small of the order of 
$ \frac{ \Gamma_{j} + \Gamma_{j'} }
       { 2\omega \left( n - n_{\mbox{\scriptsize res}} \right) } $ 
(we suppose high frequency $ \Gamma / \omega \ll 1 $). 

The resonance is caused by many - photon absorption due to the strength 
($ W \gg \omega $) of the ac-field. 
The formula (\ref{Curr-df}) expresses 
the current at the moment of time $t$ through the probability to create 
an electron in the state $m$ annihilating it in the state $k$ 
as it is drawn in fig.~\ref{res-expl}a. 
To do that it is necessary to have an electron in the state $k$ and a hole 
in the state $m$. The formula (\ref{curr-gen}) shows that it is provided 
by creation of an electron in the state $k$ and a hole in the state $n$ at 
an earlier moment of time $t_{1}, t_{1} \leq t $ (see fig.~\ref{res-expl}b). 
In the presence of a strong ac - field 
the process in which $ n \neq m $ should be considered. 
The hole then has to propagate from $n$ to $m$. 
The formula (\ref{eq-centr-<}) expresses the transition from $n$ to $m$ 
through the kind of process drawn in fig.~\ref{res-expl}c for 
$t_{3} < t_{2}$. An electron is transfered at the moment $t_{3}$, 
$t_{3} \leq t $ from the state $m$ to the state $p$ in a lead leaving a hole 
in the state $m$. At the moment $t_{2}, t_{2} \leq t_{1}$ the electron 
tunnels to 
the state $n$. Eventually an electron is found in the state $n$ at $t_{1}$ 
and a hole is found in the state $m$ at $t$. In the absence of a 
time - dependent potential this process is impossible because 
the transitions from $m$ to $p$ and from $p$ to $n$ have to conserve energy 
(after integration over $t_{2}$, $t_{3}$). Thus $m$ has to be equal to $n$. 
$H_{ac}(t)$ leads to a multiplication 
the wave function of the state $p$ by the 
phase factor $ e^{ i \frac{W_{R}}{\omega} \cos (\omega t) } $ (see 
Appendix~A and formula (\ref{leadsGF}) in the main text). In the energy 
representation the wave function becomes dispersed over all energies 
$ \epsilon_{p} \pm s \omega $, $s$ integer with the weight 
$ J_{s}(\frac{W_{R}}{\omega}) $. In other words 
the spectral function of the state $p$ has extra peaks with energies 
$ \epsilon_{p} \pm s \omega $. Their magnitudes are proportional to 
$ J_{s}^{2}(\frac{W_{R}}{\omega}) $ (see \cite{TienGordon63}). 
Thus it is possible 
to have transitions from $m$ to $p$ and from $p$ to $n$ when 
\beq
  \epsilon_{m} - \epsilon_{p} = q \omega , \hspace{5mm}    \label{res-cond}
  \epsilon_{n} - \epsilon_{p} = s \omega   
\eneq
This is the origin of 
$J_{s}$ and $J_{q}$ in the formulas (\ref{GF-sol-<}), (\ref{curr-fnl}). 
These transitions are 
accompanied by emission (absorption) of $q$~($s$) photons. The total number 
of absorbed photons is $ s - q = n $. When ac-field is strong 
($ W_{R} \gg \omega $ ) $J_{s}(\frac{W_{R}}{\omega})$ and 
$J_{q}(\frac{W_{R}}{\omega})$ can be large for $ s,q \gg 1 $. 
This leads to a possibility of many - photon absorption (emission). 
The resonance conditions (\ref{res-cond}) give 
$ n_{\mbox{\scriptsize res}} = s - q = 
	    \frac{ \epsilon_{m} - \epsilon_{n} }{ \omega } $. 
It is easy to show that, in fact, 
$ W_{R} > n_{\mbox{\scriptsize res}} \omega $ is required to obtain large 
$n_{\mbox{\scriptsize res}}$-th harmonic. 
The physical content of fig.~\ref{res-expl}c 
expresses the contribution of $Y_{nm}^{R}(t_{2},t_{3})$ into 
(\ref{eq-centr-<}). The same process but through the left lead expresses 
the contribution of  $Y_{nm}^{L}(t_{2},t_{3})$. 
At $ t_{2} < t_{3} $ the picture is very similar. The process just starts 
from a transfer of an electron from $p$ to $n$ at $t_{2}$. Then the hole 
tunnels from $p$ to $m$ at $t_{3}$. 

If we do not assume that the lead energy bands are wide (in particular 
that $\left| V_{L(R)} \right| $ is large), 
then the only change at $W_{L} = 0$ 
(which is required in order to observe 
the high frequency resonance as we show in 
the next section) is that now $ G_{n m}^{r(a)}, n \neq m $ is 
not negligible. These non-diagonal elements of the retarded
Green functions could make a positive or negative contribution 
to the process drawn in fig.~\ref{res-expl}c. Yet, we argue that the 
resonance would persist also in that case. 

Every pair of energy levels in the central region gives an independent 
contribution to the current (\ref{curr-fnl}). In the next section we consider 
one 
pair of energy levels and explore the dependence of the current it produces 
on the parameters of the system. 

\section{High harmonics generation.}
\newcommand{\nres}{\mbox{$ n_{\mbox{\scriptsize res}} $}}
\newcommand{\eo}{\epsilon_{1}}
\newcommand{\et}{\epsilon_{2}}

In this section we consider a strong ac - field ($W_{L} \gg \omega$ or 
$W_{R} \gg \omega$) and show the dependence of the current on the parameters
of the system. Since every pair of energy levels gives a separate 
contribution to the current it is enough to consider only one pair. 

Generally a strong ac - field leads to generation of high harmonics. 
In fig.\ref{broad} we plot the amplitude of the harmonics 
($ \left| I_{L}^{n} \right| $) 
versus their number $n$ for (as an example) $W_{L}=W_{R} = 30 $, 
$\eo = 11.5$, $\et = 30.5$, 
$\mu_{R} = 20$, $\mu_{L} = 5$, $V_{R} = V_{L} = -50$, 
$\Gamma_{1 1}^{L(R)} = \Gamma_{2 2}^{L(R)} = \Gamma_{1 2}^{L} = 
-\Gamma_{1 2}^{R} = 0.05 $, $\omega = 1$. 
We did not show the dc - component because its magnitude is much larger. 
We did not show the first harmonic (whose amplitude is also a few times 
larger than that of the others) since it is anyhow only a part of the 
current response with the frequency $\omega$. The other part is given by 
the capacitive currents (first terms in the formula (\ref{lead_curr}). 
The main contribution to the broadened spectrum plotted in 
fig.\ref{broad} is due to the direct 
photon - assisted transitions from the left lead to the central region and 
back described by the first term 
in the braces in $A_{j}(n)$ (see formula (\ref{curr-fnl})). 
The terms with $j \neq j'$ arising from the presence of two or more 
levels together can lead to generation of one solitary (and high) harmonic 
(see fig.\ref{one-harmonic}). In the rest of this section we describe the 
conditions required to observe one high harmonic alone and show its dependence 
on the parameters of the system. 

Since the first term in $A_{j}(n)$ 
generates many harmonics it must be eliminated. 
The easiest way to do it is to apply ac - voltage only to the right barrier: 
$W_{L}=0$ (in fact it is enough to have 
$ W_{L} \ll \left|\epsilon_{2} - \epsilon_{1} \right| $). Then 
$ A_{j}(n) = 0 , \; n\neq 0 $ because 
$ J_{s}\left(\frac{W_{L}}{\omega}\right) = J_{s}(0) = 0 , \; s\neq 0 $. 

In fig.\ref{one-harmonic} we plotted the amplitude of the harmonics 
($ \left| I_{L}^{n} \right| $) 
versus their number $n$ for $W_{L}=0$, $W_{R} = 30 $, 
$\eo = 11.5$, $\et = 30.5$, 
$\mu_{R} = 20$, $\mu_{L} = 5$, $V_{R} = V_{L} = -50$, 
$\Gamma_{1 1}^{L(R)} = \Gamma_{2 2}^{L(R)} = \Gamma_{1 2}^{L} = 
-\Gamma_{1 2}^{R} = 0.05 $, $\omega = 1$. 
As before we did not plot here the 
the dc - current and the first 
harmonic. They are are of course large compared with higher harmonics.
We notice however that if $W_{R}$ becomes larger 
than the Fermi energy in the right lead ($\mu_{R} - V_{R}$) 
they are substantially reduced, 
and the amplitude of the resonant harmonic might be comparable to the 
magnitude of the dc - current. It is clearly seen in this figure 
that among the high harmonics only the 19-th one 
($ \nres = \frac{\et-\eo}{\omega} = 19 $) is generated. 

At $W_{L}=0$ and $\Gamma_{1} \approx \Gamma_{2}$ a simpler expression 
is easily obtained for the resonant harmonic: 
\newcommand{\snres}{\mbox{\footnotesize $n_{\mbox{\tiny res}}$ }}
\beq 
   I_{L}^{\nres}  \approx 
			  \label{res-harm}
    \frac{2e}{\hbar} \left[ B_{1}(\nres) + B_{2}(\nres) \right] 
   \approx 
    i^{-\snres} \frac{2e}{\hbar} 
      \frac{ 2 \Gamma_{1 2}^{L} \Gamma_{2 1}^{R} }{\Gamma_{1} + \Gamma_{2} } 
      \sum_{ s = \frac{\eo - \mu_{R}}{\omega} }
	   ^{ \frac{\eo - V_{R}}{\omega} }  
	    J_{s}(\frac{W_{R}}{\omega}) J_{s+\snres}(\frac{W_{R}}{\omega}) 
\eneq                 
Only the terms with $s$ so that $\eo - s \omega$ is inside the right band: 
$ V_{R} < \eo - s \omega < \mu_{R} $ contribute to the sum. We notice that 
$J_{s}(\frac{W_{R}}{\omega})$ tends to zero when $s$ becomes larger than 
$\frac{W_{R}}{\omega}$ if $\frac{W_{R}}{\omega} \gg 1 $ so $V_{R}$ does not 
influence the results as soon as $ V_{R} < \eo - W_{R} $. 

In fig.\ref{bias-ac} we  drew the dependence of the resonant harmonic 
$\left( \left| I_{L}^{\nres} \right| \right) $ on both $\frac{W_{R}}{\omega}$ 
(i.e. ac - voltage) and $\frac{\eo - \mu_{R}}{\omega}$ (determined by the 
dc - bias) for $\nres = \frac{\et -\eo }{\omega} = 8 $, 
$\Gamma_{1 1}^{L(R)}/\omega = \Gamma_{2 2}^{L(R)}/\omega = 
\Gamma_{1 2}^{L}/\omega = -\Gamma_{2 1}^{R}/\omega = 0.05 $, 
$\mu_{R}/ \omega =20$, $V_{R}/ \omega = -200$. 
There is no generation ($I_{L}^{\nres} = 0$) 
if $ W_{R} < \nres \omega /2 = (\et -\eo )/2 $ because the conditions 
$|s| < \frac{W_{R}}{\omega}$ and $|s + \nres | < \frac{W_{R}}{\omega}$ 
can not be satisfied together so one of $J_{s}(\frac{W_{R}}{\omega})$, 
$J_{s+\snres}(\frac{W_{R}}{\omega})$ is small. The generation becomes 
significant for $ W_{R} > \et -\eo $. It is especially important if there 
are more than two energy levels: distant pairs of levels do not generate 
harmonics. 

The dependence on the bias (i.e. $\frac{\eo - \mu_{R}}{\omega}$) is 
oscillating. The reason is the interference of different components of the 
wave function (remember that in a strong ac - field it is spread over 
a set of energies $\ep \pm s \omega$ with different $s$). 
It shows in the formulas as oscillating behavior of 
$J_{s}(\frac{W_{R}}{\omega})$ via $s$. Let us consider the sum 
in (\ref{res-harm}) at $|\eo - \mu_{R}| \ll W_{R}$ and $V_{R} < \eo - W_{R}$. 
It is useful to divide it into two parts:
\[
    \sum_{ s = s_{1} }^{ \infty }  
	 J_{s}(\frac{W_{R}}{\omega}) J_{s+\snres}(\frac{W_{R}}{\omega})  = 
    \sum_{ s = s_{1} }^{ s^{*} }  
	 J_{s}(\frac{W_{R}}{\omega}) J_{s+\snres}(\frac{W_{R}}{\omega})  +
    \sum_{ s = s_{*} }^{ \infty }  
	 J_{s}(\frac{W_{R}}{\omega}) J_{s+\snres}(\frac{W_{R}}{\omega}) 
\]
where $s_{1} \equiv \frac{\eo - \mu_{R}}{\omega}$ and $|s^{*}| \ll W_{R}$. 
The second part does not depend on $\eo - \mu_{R}$. The Bessel's functions 
in the first part can be approximated by $\cos$: 
\[  J_{s}(\frac{W_{R}}{\omega}) = 
    \sqrt{ \frac{2}{\pi \nu} } 
    \cos \left( 
      \nu - s \arcsin \frac{\nu}{ W_{R}/\omega }  - \frac{\pi}{4} \right), \; 
    \nu \equiv \sqrt{ \left( \frac{W_{R}}{\omega} \right)^{2} - s^{2} } 
\] 
since $|s| \ll \frac{W_{R}}{\omega}$. Then 
\bea
   I_{L}^{\nres} & \approx &                             \label{curr-sin}
    i^{-\snres} \frac{2e}{\hbar} 
    \frac{ 2 \Gamma_{1 2}^{L} \Gamma_{2 1}^{R} }{\Gamma_{1} + \Gamma_{2} } 
    \frac{ 1 }{\pi W_{R}/\omega } 
    \sum_{ s = s_{1} }^{ s = s^{*} }  
       \left[ \cos \left(\alpha - \frac{\nres \omega}{W_{R}} s \right) + 
	      \cos (\beta - \pi s )                \right] + \mbox{const} 
							      \nonumber   \\ 
		 & \approx & 
    i^{-\snres} \frac{2e}{\hbar} 
    \frac{ 2 \Gamma_{1 2}^{L} \Gamma_{2 1}^{R} }{\Gamma_{1} + \Gamma_{2} } 
    \frac{ 1 }{\pi \nres } 
    \sin \left( \alpha - \nres \frac{ \eo - \mu_{R} }{W_{R}} s \right) 
				 \hspace{7mm} (\; + \; \mbox{const} )   
\enea
where $\alpha \approx \frac{\omega \snres ^{2} }{2 W_{R}} + 
		      \frac{\pi \snres }{2}                 $ 
and $ \beta \approx 2 \frac{W_{R}}{\omega} - \frac{\pi (\snres + 1)}{2} $. 
When $\eo - \mu_{R}$ changes from $-W_{R}$ to $W_{R}$ the value 
$I_{L}^{\nres}$ oscillates about $\nres / \pi $ times. The amplitude of 
the current $\left| I_{L}^{\nres} \right| $ has about $2 \nres /\pi $ 
maxima (in fig.\ref{bias-ac} the number of maxima is even larger because the 
frequency of oscillations is larger at $\eo - \mu_{R} \approx \pm W_{R} $). 
Notice that the current is not generated if $|\eo - \mu_{R}| > W_{R} $. 
It is significant for systems with many energy levels: only those pairs 
of levels that are in the energy range from $\mu_{R} - W_{R}$ to 
$\mu_{R} + W_{R}$ generate harmonics. 

The dependence of the current on the transparency of the barriers is clear 
from formulas (\ref{curr-fnl}), (\ref{res-harm}). 
In fig.\ref{one-harmonic} we showed the spectrum of the tunneling current at 
$\Gamma /\omega = 1/20$. 
If $\Gamma$ increases two sets of harmonics grow in the vicinity of  
the 0-th and the \nres~-th harmonics. 
The ratio of the amplitudes of the side harmonics to the amplitude of 
the leading (0-th or \nres~-th) one is about $\Gamma / n \omega$ and 
$\Gamma / (n -\nres) \omega$ respectively. On the other hand, the 
magnitude of the resonant current is proportional to 
$ \frac{ 2 \Gamma_{1 2}^{L} \Gamma_{2 1}^{R} } 
       { \Gamma_{1} + \Gamma_{2} } $.   

The dependence of the current on 
frequency is oscillatory. If a rectifying 
device like a diode is placed in the output the harmonics contribute to the 
dc - current. In fig.\ref{freq-depend} we draw the value 
$ \left\langle I \right\rangle \equiv 
  \frac{1}{2} \left|I_{L}^{0} \right| + 
  \frac{2}{\pi} \sum_{n=1}^{+\infty} \left|I_{L}^{n} \right| $ 
coming from time - averaging of $I_{L}^{T}(t)$ as a function of the input 
frequency $\omega$ at $W_{L}=0$, $W_{R} = 30 $, 
$\eo = 11.5$, $\et = 31.5$, 
$\mu_{R} = 20$, $\mu_{L} = 5$, $V_{R} = V_{L} = -200$, 
$\Gamma_{1 1}^{L(R)} = \Gamma_{2 2}^{L(R)} = \Gamma_{1 2}^{L} = 
-\Gamma_{1 2}^{R} = 0.05 $. The peaks correspond to different harmonics 
($ n = 22, 21, 20, 19, 18 $) satisfying the resonant condition 
$ n = \frac{\et - \eo}{\omega} $. 

We have discussed the high harmonic resonance in the tunneling current 
through the left barrier. To observe it the measuring device must be made 
sensitive to the tunneling current only through this barrier. Otherwise 
other harmonics will mask the resonance (like in fig.\ref{broad}). If the 
device measures the current in the leads we propose two ways to obtain it 
based on the formula (\ref{lead_curr}): 
\begin{enumerate} 
\item Making the gate capacitance small: $C_{g} \ll C_{L}, C_{R}$ and the 
capacitance of the right barrier much larger than the capacitance of the 
left one: $C_{L} \ll C_{R}$. It can be achieved by making the left barrier 
thicker than the right one. Notice that the right barrier must be made  
slightly higher so that $\Gamma_{L}$ and $\Gamma_{R}$ are of the same order 
(then $I_{L}^{T}$ and $I_{R}^{T}$ are of the same order). Thus the 
high harmonic currents 
in the leads are approximately given by: 
\begin{eqnarray*}
   I_{L}(t) & \approx & I_{L}^{T}(t) - \frac{C_{L}}{C_{R}} I_{R}^{T}(t) 
	      \approx   I_{L}^{T}(t)   \\
   I_{R}(t) & \approx & -I_{L}^{T}(t) + \frac{C_{L}}{C_{R}} I_{R}^{T}(t) 
	      \approx   -I_{L}^{T}(t)
\end{eqnarray*}
We omitted here the contribution of the capacitive currents having 
frequency $\omega$. 
The high harmonic currents in both leads are determined by $I_{L}^{T}$. 
\item Making the gate capacitance large: $C_{g} \gg C_{L}$ (the ratio of 
barrier capacitances $C_{L}$ and $C_{R}$ is arbitrary, there is no need to 
make one of the barriers higher than the other one). Then (dropping out the 
capacitive currents) we obtain from (\ref{lead_curr}):
\begin{eqnarray*}
   I_{L}(t) & \approx &  
		  I_{L}^{T}(t) - \frac{C_{L}}{C_{R}+C_{g}} I_{R}^{T}(t)  \\
   I_{R}(t) & \approx & 
      - \frac{C_{R}}{C_{R}+C_{g}} I_{L}^{T}(t) 
       + \frac{C_{g}}{C_{R}+C_{g}} I_{R}^{T}(t)   \nonumber 
\end{eqnarray*}
We notice that in this case high harmonics in the left lead are produced        
by $I_{L}^{T}$ but in the right lead by both $I_{L}^{T}$ and $I_{R}^{T}$ 
(unless $C_{R} \gg C_{g}$). 
If $W_{L} = 0$, $W_{R} \neq 0$ the term $A$ generating a wide spectrum of 
harmonics vanishes in $I_{L}^{T}(t)$ but it is not zero in the expression 
for $I_{R}^{T}(t)$. Then the current in the right lead $I_{R}(t)$ consists 
of many harmonics. To detect a single harmonic 
the current in the left lead should be measured. 
\end{enumerate} 

\section{Summary}
The basic physical problem addressed in this work is concerned with
a non-linear response of a two(or~more)-level system to a high frequency 
external field. 
This makes the formalism developed above particularly attractive
since two-level systems are an important model for many realistic
physical situations. The most familiar one is evidently a two-level
atomic system, whose response to a laser field 
is one of the hall marks of
non-linear quantum optics, one of whose signatures is higher frequency
generation. Here we have focused on an electronic analog, where the
response is a tunneling current instead of an emitted light. 
Note that, unlike the optical analog, the theoretical formulation requires
computation of non-equilibrium (Schwinger-Keldysh) Green functions. 
As far as the experimental situation is concerned,
we see no real obstacle in the road for
actual observation of this effect.

\noindent
We expect new effects to occur when the many-body physics is 
included. When the size of the dot is small enough then, 
at voltages and frequencies less then the 
Coulomb interaction energy (which is frequently the case), 
it can be considered
as an Anderson impurity, which shows a Kondo type effect if the
resonance level is deep below the Fermi level and the temperature
is low enough. Hence,
the first problem which comes into mind in this context
is a non-linear response of a magnetic impurity to a time-dependent field. 
The Kondo effect out of equilibrium has been studied recently by
several authors, but this is done primarily in the non-crossing
approximation, which is valid much above the Kondo temperature.
The formalism developed above gives the hope that also crossed
diagrams can be included (the first ones appear when a sixth order
term in the tunneling matrix elements are computed) and therefore
it can approach the physics at temperatures lower than the Kondo temperature.
If external voltages or frequencies are comparable with the 
Coulomb interaction energy mutual time-dependent resonant tunneling of 
electrons might show new physics. We suppose that this problem can be 
handled using the finite - $U$ Anderson model. 

\noindent 
We would like to thank A. Golub, J. Golub and N. S. Wingreen for very
helpful discussion. The research of Y. A is partially supported by 
the Fund for Basic Research of the Israeli Academy of Science and
by the Israeli Ministry of Science and the Arts.

\appendix 
\section*{Appendix A} 
       
In this Appendix we show that an arbitrary strong time - dependent potential 
has no effect if it is uniform in space. We also 
calculate here Green functions for an 
isolated lead. 

If the alternating potential ($H_{ac}(t)$) is uniform the Hamiltonian can be 
written in the following form: 
\beq
   H(t) = H_{f} + H_{ac}(t)                     \label{AppHam}
\eneq 
where $H_{f}$ is time - independent, $H_{ac}(t)$ is space - independent. 
The solutions of the Schr\"{o}dinger's equation are: 
\beq 
   \Psi_{k}(t) = \varphi_{k}(t) 
		e^{ -\frac{i}{\hbar} \int^{t} H_{ac}(t_{1}) dt_{1} }
\eneq        
where $\varphi_{k}(t)$ are eigenstates of $H_{f}$. 
The time dependent part $H_{ac}(t)$ 
influences only the phases giving all the solutions {\em the same\/} phase 
factor $ e^{ -\frac{i}{\hbar} \int^{t} H_{ac}(t_{1}) dt_{1} } $. The phase 
differences are determined by $H_{f}$ only. Thus $H_{ac}(t)$ has no physical 
effect. This simple result is the time - dependent analog of the fact that 
the (time - independent) reference point of energy can be arbitrarily chosen. 

We used this result to write the ac - part of the Hamiltonian of a double - 
barrier structure in the form (\ref{Ham-ac}). The ac - shift of the central 
region is ignored. 

In an isolated lead the ac - potential is uniform (see (\ref{Ham-ac})). 
Then the evolution operator 
$\left( U(t,t') \right)$ is obtained from the equation 
$ i \hbar \frac{\partial}{\partial t} U(t,t') = H(t) U(t,t') $ 
with $H(t)$ given by (\ref{AppHam}). Its matrix elements are: 
\beq 
   U_{k k'}(t,t') = \delta_{k k'}                 \label{ev-oper}
		e^{ -\frac{i}{\hbar} \int_{t'}^{t} H_{k k}(t_{1}) dt_{1} }
\eneq
$\delta_{k k'}$ appears because $H_{k k'} = 0, \; k \neq k'$ (i.e a uniform 
potential does not cause transitions). Using the definition of 
Green functions in the 
Schr\"{o}dinger representation with the evolution operator (\ref{ev-oper}) 
it is easy to obtain the formulas (\ref{leadsGF}). 

\appendix 
\section*{Appendix B} 

In this Appendix we analyze the Dyson's equation for $G_{n m}^{<}$ and 
obtain the formula (\ref{eq-centr-<}). The Dyson's equation (\ref{Dyson-<}) 
holds for $G_{n m}^{<}$: 
\bea 
  G_{n m}^{<} & = & F_{n m}^{1}(g_{j j}^{<}) + 
		    F_{n m}^{2}(g_{k k}^{<}, g_{p p}^{<})            \\ 
  F_{n m}^{1}(g_{j j}^{<}) & \equiv &  g_{n m}^{<} 
	+ G_{n k}^{r} T_{k m} g_{m m}^{<} + G_{n p}^{r} T_{p m} g_{m m}^{<} + 
							       \nonumber  \\
    & + & g_{n n}^{<} T_{n k} G_{k m}^{a} + g_{n n}^{<} T_{n p} G_{p m}^{a} +
							       \nonumber  \\
    & + &  G_{n k}^{r} T_{k i} g_{i i}^{<} T_{i k'} G_{k' m}^{a}
	 + G_{n p}^{r} T_{p i} g_{i i}^{<} T_{i p'} G_{p' m}^{a} + 
							       \nonumber  \\
    & + &  G_{n k}^{r} T_{k i} g_{i i}^{<} T_{i p} G_{p m}^{a}
	 + G_{n p}^{r} T_{p i} g_{i i}^{<} T_{i k} G_{k m}^{a} \nonumber  \\
  F_{n m}^{2}(g_{k k}^{<}, g_{p p}^{<}) & \equiv & 
	   G_{n n'}^{r} T_{n' k} g_{k k}^{<} T_{k m'} G_{m' m}^{a}
	 + G_{n n'}^{r} T_{n' p} g_{p p}^{<} T_{p m'} G_{m' m}^{a} \nonumber
\enea 
Here multiplication implies integration over time and summation over 
repeated indexes; $j=m,n,i$ belong to the central region, $k, k'$ --- to 
the left lead, $p, p'$ --- to the right lead. 

$F_{n m}^{1}(g_{j j}^{<})$ depends on the initial state of the central 
region. It does not depend on $g_{k k}^{<}, g_{p p}^{<}$ 
($G^{r}$, $G^{a}$ are determined by the equation (\ref{Dyson-r}) 
which does not depend on $g_{k k}^{<}, g_{p p}^{<}$). It describes the 
behavior of the system with empty leads. Indeed, if 
$g_{k k}^{<} = 0, g_{p p}^{<} = 0 $ we have $F_{n m}^{2} = 0 $. Then 
$ G_{n m}^{<} = F_{n m}^{1} $. When the leads are empty the central region 
empties with time. $F_{n m}^{1}$ tends to zero. After a long enough period 
of time has passed from the moment the tunneling was switched on it can be 
neglected. Then $ G_{n m}^{<} = F_{n m}^{2}$. This is the formula 
(\ref{eq-centr-<}). We can say that $F_{n m}^{1}$ describes the transient 
processes while $F_{n m}^{2}$ gives some kind of 
a quasistationary (but fully time -- dependent) solution.

\newpage
\begin{figure}[t]
\caption{Schematic energy diagram of the conduction band for an 
	 $\mbox{\em  Al}_{x}\mbox{\em  Ga}_{1-x}\mbox{\em  As}$
	 quantum well in the absence of an ac-field. $x$ is the 
	 axis perpendicular to the layers, $\mu_{L}$ and $\mu_{R}$ 
	 are the left and the right chemical potentials correspondently, 
	 $V_{L}$ and $V_{R}$ are the energies of the conduction 
	 band bottoms of the leads, $\epsilon_{1}$ and $\epsilon_{2}$ 
	 are the energies of the levels in the central region. 
	 \label{dox-1} }
\caption{ Tunneling diagrams. A cross denotes a tunneling event, a thing 
	  line -- free propagator, a thick line -- full propagator. 
	  a) Dyson's equation. Only a single tunneling event contributes to 
	  the self - energy. 
	  b) any other diagram is reducible. \label{tunn-diagrams} }
\caption{Tunneling processes leading to the high harmonic resonance. 
	 A thick black line denotes full electron propagator, a thick 
	 white one -- full hole propagator, a thing black line -- free 
	 electron propagator, a dashed line -- tunneling event. 
	 a) The tunneling current is produced by electrons going from 
	 the lead to a level in the central region and back. 
	 b) This process can be assisted by other levels if ac - field is 
	 strong. 
	 c) Resonance transfer of an electron from the level $m$ to the level 
	 $n$ possible in a strong ac-field.   \label{res-expl} }
\caption{Spectrum of the tunneling current (amplitude of the harmonics via 
	 their number). General situation. $W_{L}=W_{R} = 30 $, 
	 $\eo = 11.5$, $\et = 30.5$, 
	 $\mu_{R} = 20$, $\mu_{L} = 5$, $V_{R} = V_{L} = -50$, 
	 $\Gamma \equiv 
	  \Gamma_{1 1}^{L(R)} = \Gamma_{2 2}^{L(R)} = \Gamma_{1 2}^{L} = 
	  -\Gamma_{1 2}^{R} = 0.05 $, $\omega = 1$. 
	 The dc - current and the 1-st harmonic are not shown. 
	 \label{broad} }
\caption{Spectrum of the tunneling current (amplitude of the harmonics via 
	 their number) when the alternating field is applied on the right  
	 barrier: $W_{L}=0$, $W_{R} = 30 $, $\eo = 11.5$, $\et = 30.5$, 
	 $\mu_{R} = 20$, $\mu_{L} = 5$, $V_{R} = V_{L} = -50$, 
	 $\Gamma \equiv 
	  \Gamma_{1 1}^{L(R)} = \Gamma_{2 2}^{L(R)} = \Gamma_{1 2}^{L} = 
	 -\Gamma_{1 2}^{R} = 0.05 $, $\omega = 1$. 
	 The dc - current and the 1-st harmonic are not shown. 
	 \label{one-harmonic} }
\caption{Dependence of the resonant harmonic 
	 $\left( \left| I_{L}^{\nres} \right| \right) $ on the dc - bias 
	 (i.e. $\eo - \mu_{R}$) and the amplitude of the ac - voltage 
	 $W_{R}$. $\nres = 8 $, $W_{L}=0$, $\et = \eo + 8$, 
	 $\mu_{R} = 20$, $V_{R} = -200$, 
	 $\Gamma_{1 1}^{L(R)} = \Gamma_{2 2}^{L(R)} = \Gamma_{1 2}^{L} = 
	 -\Gamma_{1 2}^{R} = 0.05 $, $\omega = 1$. 
	 \label{bias-ac} }
\caption{The output current rectified by a diode versus the frequency 
	 $\omega$ of the input ac - voltage. 
	 $W_{L}=0$, $W_{R} = 30 $, $\eo = 11.5$, $\et = 31.5$, 
	 $\mu_{R} = 20$, $\mu_{L} = 5$, $V_{R} = V_{L} = -200$, 
	 $\Gamma \equiv 
          \Gamma_{1 1}^{L(R)} = \Gamma_{2 2}^{L(R)} = \Gamma_{1 2}^{L} = 
	 -\Gamma_{1 2}^{R} = 0.05 $. The peaks correspond to different 
	 harmonics ($ n = 22, 21, 20, 19, 18 $) satisfying the resonant 
	 condition $ n = \frac{\et - \eo}{\omega} $. 
	 \label{freq-depend} }
\end{figure} 


\begin{thebibliography}{ JonsonGrincwaig87}
\fussy
\bibitem{Price94} J. B. Pieper and J. C. Price, 
Phys. Rev. Lett. {\bf 72}, 3586 (1994).
  \bibitem{ButtPrTh93} M.B\"{u} ttiker, A.Pr\^{e}tre, H.Thomas,
		      Phys.Rev.Lett., 70, 4114 (June~1993).
\bibitem{Liu94} D. Z. Liu, B. Y. K. Hu, C. A. Stafford and S. Das Sarma,
Phys. Rev. {\bf B50}, 5799 (1994).
\bibitem{Pieper94} J. B. Pieper and J. C. Price, 
Phys. Rev. {\bf B49}, 17059 (1994).
\bibitem{Greelings90} L. J. Greelings {\it et al.}, Phys. Rev. Lett.
{\bf 64}, 2691 (1990).
\bibitem{Kouwenhoven91} L.P.Kouwenhoven, A.T.Johnson, 
		 N.C.van~der~Vaart, C.J.P.M.Harmans, C.T.Foxon,
		 Phys.Rev.Lett., 67, 1626 (1991).
\bibitem{Wingreen96} C.A.Stafford, N.D.Wingreen, 
{\it Resonant Photon-Assisted Tunneling Through a
Quantum Dot: An Electron Pump from Spatial Rabi Oscillations},
preprint VGVA-DPT 1995/09-901
\bibitem{Faist94} J. Faist {\it et al.}, Science, {\bf 264}, 553 (1994).
  \bibitem{TienGordon63} R.K.Tien, J.P.Gordon, Phys.Rev., 129, 647 (1963).
  \bibitem{KouwMcEuen-PRB} L.P.Kouwenhoven, S.Jauhar, K.McCormick, D.Dixon, 
			   P.L.McEuen, Yu.V.Nazarov, N.C.van~der~Vaart, 
			   C.T.Foxon, Phys.Rev.B, 50, 2019 (July~1994). 
  \bibitem{KouwMcEuen-PRL} L.P.Kouwenhoven, S.Jauhar, J.Orenstein,  
			   P.L.McEuen, Y.Nagamune, J.Motohisa, H.Sakaki,  
			   Phys.Rev.Lett., 73, 3443 (Dec.~1994). 
  \bibitem{BruScho94} C.Bruder, H.Schoeller, Phys.Rev.Lett., 
		      72, 1076 (Feb.~1994). 
\bibitem{Akiyama94} H. Akiyama {\it et al.}, Appl. Phys. Lett.
{\bf 65}, 424 (1994).
\bibitem{Blick} R. H. Blick, R. J. Haug, K. von Klitzing and K. Eberl, 
preprint. 
  \bibitem{ChEsTsu74} L.L.Chang, L.Esaki, R.Tsu, Appl.Phys.Lett.,
		      24, 593 (June~1974).
  \bibitem{Sollner83} T.C.L.G.Sollner, W.D.Goodhue, P.E.Tannenwald,
		      C.D.Parker, D.D.Peck, Appl.Phys.Lett.,
		      43, 588 (Sept.~1983).
  \bibitem{Sollner84} T.C.L.G.Sollner, P.E.Tannenwald, D.D.Peck, W.D.Goodhue,
		      Appl.Phys.Lett., 45, 1319 (Dec.~1984).
  \bibitem{Whitaker88} J.F.Whitaker, G.A.Mourou, T.C.L.G.Sollner, W.D.Goodhue,
		       Appl.Phys.Lett., 53, 385 (Aug.~1988). 
  \bibitem{Goldman87-PR-L} V.J.Goldman, D.C.Tsui, J.E.Cunningham,
			   Phys.Rev.Lett., 58, 1256 (March~1987).
  \bibitem{Goldman87-PR-B} V.J.Goldman, D.C.Tsui, J.E.Cunningham,
			   Phys.Rev.B, 35, 9387 (June~1987).
  \bibitem{Reed88} M.A.Reed, J.N.Randall, R.J.Aggarwal, R.J.Matyi,
		   T.M.Moore, A.E.Wetsel, Phys.Rev.Lett.,
		   60, 535 (Feb.~1988).
  \bibitem{Rydberg89} A.Rydberg, H.Gr\"{o}nqvist, Electr.Lett.,
		      25, 348 (March~1989).
  \bibitem{Brown91} E.R.Brown, J.R.S\"{o}derstr\"{o}m, C.D.Parker,
		    L.J.Mahoney, K.M.Molvar, T.C.McGill,
		    Appl.Phys.Lett., 58, 2291 (May~1991).
  \bibitem{SuGoldCunn92} Bo~Su, V.J.Goldman, J.E.Cunningham,
			Science, v.255, p.313, Jan.~1992.  
  \bibitem{Dellow} M.W.Dellow, P.H.Beton, C.J.G.M.Langerak, T.J.Foster, 
		   P.C.Main, L.Eaves, M.Henini, S.P.Beaumont, C.D.W.Wilkinson, 
		   Phys.Rev.Lett., 68, 1754 (Mar.~1992). 
  \bibitem{Gueret} P.Gu\'{e}ret, N.Blanc, R.Germann, .Rothuizen, 
		   Phys.Rev.Lett., 68, 1896 (Mar.~1992). 
  \bibitem{Ashoori92} R.C.Ashoori, H.L.Stormer, J.S.Weiner, L.N.Pfeiffer,
	   S.J.Pearton, K.W.Baldwin, K.W.West, Phys.Rev.Lett., 68,
	   3088 (May~1992).
  \bibitem{Meirav90} U.Meirav, M.A.Kastner, S.J.Wind,
		     Phys.Rev.Lett., 65, 771 (Aug.~1990).
  \bibitem{McEuenMWK91} P.L.McEuen, E.B.Foxman, U.Meirav,
	   M.Kastner, Y.Meir, N.S.Wingreen, S.J.Wind, Phys.Rev.Lett., 66,
	   1926 (Apr.~1991). 
  \bibitem{FordPepper} C.J.B.Ford, P.J.Simpson, M.Pepper, D.Kern,
	   J.E.F.Frost, D.A.Ritchie, G.A.C.Jones, 
  \bibitem{FieldPepper93} M.Field, C.G.Smith, M.Pepper, D.A.Ritchie,
			  J.E.F.Frost, G.A.C.Jones, D.G.Hasko,
			  Phys.Rev.Lett., 70, 1311 (Mar.~1993).
  \bibitem{FoxmanMWK93} E.B.Foxman, P.L.McEuen, U.Meirav,
	   N.S.Wingreen, Y.Meir, P.A.Belk, N.R.Belk, M.Kastner,
	   Phys.Rev.B, 47, 10020 (Apr.~1993).
  \bibitem{KlitzPloog93} J.Weis, R.J.Haug, K.v.Klitzing, K.Ploog,
			Phys.Rev.Lett., 71, 4019 (Dec.~1993).
  \bibitem{MeurerPloog} B.Meurer, D.Heitmann, K.Ploog, Phys.Rev.Lett., 
			68, 1371 (Mar.~1992). 
  \bibitem{JohnKouwenh92} A.T.Johnson, L.P.Kouwenhoven, W.de~Jong,
			 N.C.van~der~Vaart, C.J.P.M.Harmans, C.T.Foxon,
			 Phys.Rev.Lett., 69, 1592 (Sept.~1992).
  \bibitem{Nakazato} K.Nakazato, T.J.Thornton, J.White, H.Ahmed, 
		     Appl.Phys.Lett., 61, 3145 (Dec.~1992). 
  \bibitem{Heiblum} A.Yacoby, M.Heiblum, D.Mahalu, H.Shtrikman, 
		    Phys.Rev.Lett., 74, 4047 (May~1995). 
  \bibitem{Fowler86} A.B.Fowler, G.L.Timp, J.J.Wainer, R.A.Webb, 
		     Phys.Rev.Lett., 57, 138 (July~1986). 
  \bibitem{Kopley88} T.E.Kopley, P.L.McEuen, R.G.Wheeler, 
		     Phys.Rev.Lett., 61, 1654 (Oct.~1988). 
  \bibitem{Molenkamp} L.W.Molenkamp, K.Flensberg, M.Kemerink, 
  \bibitem{RalphBuhr} D.C.Ralph, R.A.Buhrman, 
		      Phys.Rev.Lett., 72, 3401 (May~1994). 
  \bibitem{Bardeen} J.Bardeen, Phys.Rev.Lett., 6, 2 (Jan.~1961). 
  \bibitem{Cohen} M.H.Cohen, L.M.Falicov, J.C.Phillips, Phys.Rev.Lett., 
		  8, 316 (Apr.~1962). 
  \bibitem{ChenTing91} L.Y.Chen, C.S.Ting, Phys.Rev.B, 43, 2097 (1991).
  \bibitem{Frensley88} W.R.Frensley, Superlattices and Microstructures,
		       4, 497 (1988).
  \bibitem{MW92} Y.Meir, N.S.Wingreen, Phys.Rev.Lett., 68, 2512 (Apr.~1992).
  \bibitem{Sugimura93} A.Sugimura, Phys.Rev.B, 47, 9676 (Apr.~1993).
  \bibitem{Tucker79} J.R.Tucker, IEEE~J.Quant.Elec., QE-15, 1234 (1979).
  \bibitem{Johansson90} P.Johansson, Phys.Rev.B, 41, 9892 (1990).
  \bibitem{Liu91} H.C.Liu, Phys.Rev.B, 43, 12538 (1991).
  \bibitem{Cai91} W.Cai, P.Hu, M.Lax, Phys.Rev.B, 44, 3336 (1991).
  \bibitem{SuYu91} Z.-B.Su, Lu~Yu, L.-Y.Chen, in:
		  {\em Thermal Field Theories},
		  eds. H.Egawa, T.Arimitru, Y.Hashimoto, p.107, 
		  (Elsevier, 1991). 
  \bibitem{Isawa92} Y.Isawa,in {\em Transport Phenomena in Mesoscopic
		  Systems}, ed. by H.Fukuyama and T.Ando, Springer
		  Series in Solid-State Sciences, v.109, p.93,
		  (Springer, Berlin, 1992).
  \bibitem{Avishai} R.Berkovits, M.Abraham, Y.Avishai,
		    J.Phys.: Condensed. Matter, 5, L175 (1993).
  \bibitem{MW93:ac} N.S.Wingreen, A.P.Jauho, Y.Meir, Phys.Rev.B, 48,
		 8487 (Sept.~1993).
  \bibitem{ChenTing90} L.Y.Chen, C.S.Ting, Phys.Rev.Lett., 64, 3159 (1990). 
  \bibitem{RiccoAzbel84} B.Ricco, M.Ya.Azbel, Phys.Rev.B, 29, 1970
			 (Feb.~1984).
  \bibitem{StAzLee85} A.D.Stone, M.Ya.Azbel, P.A.Lee, Phys.Rev.B,
		      31, 1707 (1985).
  \bibitem{WeilVinter87} T.Weil, B.Vinter, Appl.Phys.Lett.,
			50, 1281 (May~1987).
  \bibitem{JonsonGrincwaig87} M.Jonson, A.Grincwaig, Appl.Phys.Lett.,
			      51, 1729 (Nov.~1987).
  \bibitem{Buttiker88} M.B\"{u}ttiker, IBM~J.Res.Develop., 32,
		      63 (Jan.~1988).
  \bibitem{Sok-JPC-88} D.Sokolovski, J.Phys.C: Solid State Phys.,
		      21, 639 (1988).
  \bibitem{SumFel88} M.Yu.Sumetskii, M.L.Fel'shtyn, Sov.Phys.JETP,
		    67, 1610 (1988).
  \bibitem{Price92} P.Price, Phys.Rev.B, 43, 2097 (Jan.~1991). 
  \bibitem{SheardToombs88} F.W.Sheard, G.A.Toombs, Appl.Phys.Lett.,
			   52, 1228 (Apr.~1988).
  \bibitem{Averin91} D.V.Averin, A.N.Korotkov, K.K.Likharev, 
		     Phys.Rev.B, 44, 6199 (Sept.~1991).
  \bibitem{FiigJauho92} T.Figg, A.P.Jauho, Surf.Sci., 267, 392 (1992). 
  \bibitem{Schwinger61} J.Schwinger, J.Math.Phys., 2, 407 (1961).
  \bibitem{KadBaym62} L.P.Kadanoff, G.Baym, {\em Quantum Statistical
		     Mechanics}, 203p., W.A.Benjamin Inc., N.Y., 1962.
  \bibitem{Keldysh65} L.V.Keldysh, Sov.Phys.JETP, 20, 1018 (1965).
  \bibitem{LandauLif81} L.D.Landau, E.M.Lifshitz, {\em Course of Theoretical 
			Physics, v.10, Physical Kinetics}, p.395, 
			(Pergamon Press, 1981). 
  \bibitem{ChouSuHaoYu85} K.-C.Chou, Z.-B.Su, B.-L.Hao, Lu~Yu, Phys.Rep.
			 (Rev. Section. Phys.Lett), 118, 1 (1985).
  \bibitem{Mahan-b-90} G.D.Mahan, {\em Many-Particle Physics}, 1990.
  \bibitem{Mahan87} G.D.Mahan, Phys.Rep. (Rev. Section. Phys.Lett), 
		   145, 251 (1987).
  \bibitem{RammerSm86} J.Rammer, H.Smith, Rev.Mod.Phys., 58, 323 (1986).
  \bibitem{CarComNozSJ-1-71} C.Caroli, R.Combescot, P.Nozieres,
			    \mbox{D.Saint-James},
			    J.Phys.C: Solid State Phys., 4, 916 (1971).
  \bibitem{Wingreen89} N.S.Wingreen, K.W.Jacobsen, J.W.Wilkins,
		      Phys.Rev.B, 40, 11834 (Dec.~1989).
  \bibitem{LangrethNor91} D.C.Langreth, P.Nordlander, Phys.Rev.B, 
			  43, 2541 (Feb.~1991). 
  \bibitem{IngoldNaz} G.-L.Ingold, Yu.V.Nazarov, in: {\em Single Charge 
		      Tunneling}, ed. by H.Grabert, M.H.Devoret, N.Y., 1992. 

\end{thebibliography}
\end{document}